\documentclass[10pt,twocolumn,twoside]{IEEEtran}

\IEEEoverridecommandlockouts                             
\usepackage[pdftex]{graphicx}
\graphicspath{{figure/}}
\usepackage{amsmath}
\usepackage{amssymb}
\usepackage{url}

\usepackage{amsthm} 
\usepackage{cite}
\usepackage{color}
\usepackage{comment}
\usepackage{fancyhdr} 

\usepackage{algorithm}
\usepackage{algorithmic}
\usepackage{bm}
\usepackage{bbm}

\newtheorem{theorem}{\textbf{Theorem}}
\newtheorem{lemma}{\textbf{Lemma}}
\newtheorem{assumption}{\textbf{Assumption}}

\newcommand{\bx}{\bm{x}}
\newcommand{\bth}{\bm{\theta}}
\newcommand{\bTh}{\bm{\Theta}}
\newcommand{\by}{\bm{y}}

\newcommand{\R}{\mathbb{R}}

\newcommand{\sS}{\mathcal{S}}
\newcommand{\sP}{\mathcal{P}}
\newcommand{\sN}{\mathcal{N}}

\newcommand{\F}{\mathcal{F}}

\newcommand{\E}{\mathbb{E}}

\newcommand{\bmu}{\bm{\mu}}

\allowdisplaybreaks

\title{Online Residential Demand Response via Contextual Multi-Armed Bandits}

\author{Xin Chen, Yutong Nie, Na Li
\thanks{X. Chen and N. Li are with the School of Engineering and Applied Sciences, Harvard University, USA; emails: chen\_xin@g.harvard.edu, nali@seas.harvard.edu. Y. Nie is with the School of Mathematical Sciences, Zhejiang University, China; email: ytnie@zju.edu.cn.}
\thanks{ 
The work was supported by
NSF 1608509, NSF CAREER 1553407, and AFOSR YIP.} 
}

\begin{document}

\maketitle

	
\begin{abstract}

Residential loads have a great potential to enhance the efficiency and reliability of electricity systems via demand response (DR) programs. One major challenge in residential DR is to  handle the unknown and uncertain customer behaviors.  Previous works  use learning techniques to  predict customer DR behaviors, while the influence of time-varying environmental factors are generally neglected, which may lead to inaccurate prediction and inefficient  load adjustment.
In this paper, we consider the residential DR problem where the load service entity (LSE) aims to select an optimal subset of customers to maximize the expected load reduction with a financial budget. 
To learn the uncertain customer behaviors under  environmental influence,  we formulate the residential DR  as a  contextual multi-armed bandit (MAB)  problem, and the online learning and selection (OLS) algorithm  based on Thompson sampling  is proposed to solve it. This algorithm takes the contextual information into consideration and is applicable to complicated DR settings. 
Numerical simulations are  performed to demonstrate the learning effectiveness of the proposed algorithm.

\end{abstract}


\section{Introduction} \label{sec:introduction}

\IEEEPARstart{W}{ith}  the deepening penetration of renewable generation and growing peak load demands, power systems 
are inclined to confront a deficiency of reserve capacity for power balance. Instead of deploying additional generators, demand response (DR) is an economical and environmentally friendly  alternative solution that
motivates the change of flexible load demands to fit the needs of power supply. 
Most of the existing DR programs focus on industrial and large commercial consumers through the direct load control and interruptible loads \cite{ DRprog2,DRprog3}. While  residential loads actually take up a significant share of the total electricity usage (e.g. about 38\%   in the U.S. \cite{eleuseUS}),
 which have a  huge potential to be exploited to facilitate  
power system operation. Moreover, the proliferation of smart meters,  smart sensors, and automatic devices, enables the remote  monitoring and control of  home electric appliances with a two-way communication between load service entities (LSEs) and  households. As a consequence, residential DR has attracted a great deal of recent interest from both academia and industry.

The mechanisms for residential DR are mainly categorized as price-based  and incentive-based \cite{drreview}. The price-based DR programs \cite{price1,price2,price3} use various pricing schemes, such as time-of-use pricing, critical peak pricing, and real-time pricing, to influence and guide the residential electricity usage. While the reactions of customers to the price change signals are highly uncontrollable and may lead to undesired load dynamics like rebound peak \cite{price1}. In  incentive-based DR programs  \cite{ince1, ince2,ince3}, the LSEs
recruit customers to participate in an upcoming DR event with financial incentives, e.g. cash, coupon, raffle, rebate, etc. During the DR event, 
customers  reduce their electricity consumption to earn the revenue but are allowed to opt out at any time. For the LSEs, it is significant to target  appropriate customers among the large population, since
each recruitment comes with a cost. Accordingly,
this paper focuses  on the incentive-based residential DR programs, and studies the strategies to select right customers  
for load reduction
from the perspective of LSEs.

In practical DR implementation, the LSEs confront a major challenge that the customer behaviors to the incentives are uncertain and unknown. 
According to the investigations in \cite{survey1, survey2, survey3},  
the user  acceptances of DR load control
are influenced by individual preference and environmental factors.
The former  relates to customers' intrinsic socio-demographic characteristics, e.g., income, age, education, household size, attitude to energy saving, etc. The latter refers to immediate externalities such as 
indoor temperature,  offered revenue, electricity price, fatigue effect, weather conditions,  etc. However,  LSEs barely  have the access to customers' individual preferences or the knowledge how  environmental factors affect their opt-out behaviors. Without considering the actual willingness, a blind customer selection scheme for DR participation  may lead to high opt-out rate and inefficient load shedding.

To address this challenge, 
a natural idea is to learn unknown customer behaviors through interaction and observation. In particular,  the multi-armed bandit (MAB) framework \cite{mab1} can be used to model the residential DR as an online learning and decision-making problem. 
MAB deals with the uncertain optimization problems
where an agent selects actions (arms) sequentially and upon each action observes a reward, while the agent is  uncertain about how his actions influence the rewards. 
After each action-reward pair is observed, the agent learns about the rewarding system,  and this allows him to improve the action strategies. In addition,
 when the reward (or outcome) to each action is not fixed but affected by some contexts, e.g. environmental factors and agent profiles,
such problems are referred as contextual MAB (CMAB) \cite{mab1}. 
Due to
its useful structure, (C)MAB  has been successfully applied in many fields, such as 
recommendation system \cite{recsys}, clinical trial \cite{clin}, web advertisement \cite{adver}, and etc.

To achieve high performance in MAB problems, it necessitates  an effective balance between {exploration} and {exploitation}, i.e., whether taking a risk to explore  poorly understood actions or exploiting current knowledge for decision-making.
To this end,  upper confidence bound (UCB) and Thompson sampling (TS) are two prominent algorithms that solve MAB problems.
UCB algorithms \cite{ucb1,ucb2} employ the upper confidence bound as an optimistic estimate of the mean reward for each action and encourage the exploration with optimism. In a Bayesian style, TS algorithms \cite{ts1} form a prior distribution over the unknown parameters and select  actions based on a random sample from the posterior distribution, which motivate the exploration via sampling randomness. Without the necessity of sophisticated design for UCB functions, 
TS has a simple and flexible solution structure that can be generalized to complex online optimization problems, 
and generally achieves better empirical performance than UCB algorithms \cite{ts2}.

For the residential DR, most existing literature does not  consider or overly simplify the uncertainty of customer behaviors, which  causes a disconnection between theory and practice. A number of studies \cite{rl1,rl2,primab,onpri,online5,online6,online7,online8} use online learning techniques to deal with the unknown information in DR problems. In terms of the price-based DR, references \cite{rl1,rl2} use reinforcement learning to learn the customer dissatisfaction on job delay, and automatically schedule the household electricity usage
under the time-varying price. From the perspective of LSEs,
references \cite{primab} and \cite{onpri} design the dynamical pricing schemes based on the MAB method and an online linear regression algorithm respectively, considering the uncertain user responses to the price change. For the incentive-based DR,  in \cite{online5, online6,online7,online8}, 
MAB and its variations like adversarial MAB, restless MAB, are utilized to learn customer behaviors and unknown load parameters and select right customers to send DR signals, where  UCB algorithms, policy index, and other heuristic algorithms are applied for solution. In these researches, simplified DR problem formulations are employed, and the effects of time-varying environmental factors on customer behaviors are mostly  neglected.

In this paper, to deal with the uncertain customer behaviors and the environmental influence, we adopt the contextual MAB to model the residential DR as an online learning and decision-making problem.  {
Specifically, this paper studies the  DR problem where the LSE aims to select right customers to maximize the total expected load reduction under a certain budget. Based on the Thompson sampling (TS) framework \cite{ts1}, we develop 
an online learning and (customer) selection (OLS) algorithm to tackle this problem, where the logistic regression  \cite{logis} is used to predict  customer opt-out behaviors and the variational approximation approach \cite{varia} is employed for efficient Bayesian inference. 
With the decomposition into a Bayesian learning task and an offline optimization task,  the proposed  algorithm is applicable to practical DR applications with complicated  settings. 
The main contribution of this paper is that  the contextual information, including individual diversity and time-varying environmental factors, is taken into consideration in the learning process, which can improve the predictions on customer DR behaviors and lead to efficient customer selection schemes.}

Besides, the theoretical performance guarantee on the proposed algorithm is provided, which shows that a sublinear Bayesian regret can be achieved.

The remainder of this paper is organized as follows: Section \ref{sec:problem} introduces the  residential DR problem formulation with uncertainty. Section \ref{sec:algorithm} develops the  online learning and customer selection algorithm. 
 Section \ref{sec:regret} analyzes the performance of the proposed OLS algorithm.
  Numerical tests are carried out in Section \ref{sec:simulation}, and conclusions are drawn in Section \ref{sec:conclusion}.

\section{Problem Formulation} \label{sec:problem}


\subsection{Residential DR Model}

\begin{figure}[thpb]
	\centering
	\includegraphics[scale=0.6]{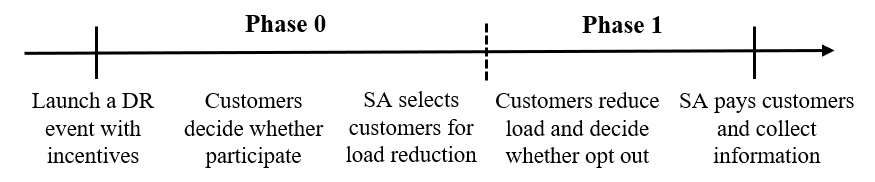}
	\caption{Two phases of a residential DR event.}
	\label{fig:twophase}
\end{figure}

Consider  the residential DR program with a system aggregator (SA) and $N$  customers over a time horizon $[T]:=\{1,\cdots,T\}$, where each time  $t\in[T]$ corresponds to one DR event. As illustrated in Figure \ref{fig:twophase}, there are two phases in a typical DR event \cite{drmech}. Phase 0 denotes the preparation period, when the  SA calls upon  customers for load reduction  with incentives and selects a subset of participating customers under a certain budget. In phase 1, the selected customers decrease their electric usage, e.g., shut down the air conditioner or increase the setting temperature, while they can choose to opt out if unsatisfied.  In the end, the SA pays the selected customers according to their contributions to the load reduction.

For customer $i\in [N]:=\{1,\cdots,N\}$, let $d_{i,t}$ be the load demand that can be shut off at $t$-th DR event, and $r_{i,t}$ be the   revenue (or the bidding price) for such load reduction. 
Denote $z_{i,t}\in\{1,0\}$ as the binary variable indicating whether customer $i$  stays in (equal $1$) or opts out (equal $0$) during $t$-th DR event if  selected. 
Assume that  $z_{i,t}$ is a random variable 
following Bernoulli distribution with $z_{i,t}\sim \text{Bern}(p_{i,t})$, which is independent across times and customers.  Without loss of generality, assume that all customers decide to participate in every DR event at phase 0, otherwise let  $r_{i,t}$ be sufficiently large 
for the unparticipated customers.

This paper studies the customer selection strategies at phase 0 from the perspective of the SA.
At $t$-th DR event, based on the reported $(d_{i,t}, r_{i,t})_{i\in[N]}$,   the SA aims to select a subset of customers to achieve certain DR goals under the  given budget $b_t$. 
Accordingly, the optimal customer selection (OCS) problem  can be formulated as 
\begin{subequations} \label{eq:ocs}
	\begin{align}
\text{Obj.} \  &	\max_{\mathcal{S}_t \subseteq [N]}\, g_t\left( (z_{i,t})_{i\in[N]},(d_{i,t})_{i\in[N]}, \mathcal{S}_t\right) \\ \ 
\text{s.t.} \ & \  h_t\left( (z_{i,t})_{i\in[N]}, (r_{i,t})_{i\in[N]},\mathcal{S}_t\right) \leq b_t, \ \mathcal{S}_t \in \Pi_t\label{eq:ocs:con} 
	\end{align}
\end{subequations}
where $g_t$ and $h_t$ represent the objective function and the cost function respectively. $\mathcal{S}_t$ is  the decision variable denoting the set of selected customers. $\Pi_t$ is the feasible set of $\mathcal{S}_t$ that describes other physical constraints. For example,  the network and power flow constraints can be captured by $\Pi_t$, which is elaborated in Appendix \ref{app:powerflow}.  

The OCS model (\ref{eq:ocs}) is a general formulation. Depending on  practical DR applications, functions $g_t$ and  $h_t$ can take different forms. Two concrete examples are provided as follows.

\textbf{Example 1.} Model (\ref{eq:exam1}) maximizes the total expected load reduction, and  constraint (\ref{eq:exam1:con}) ensures that the revenue payment does not exceed the given budget.
\begin{subequations} \label{eq:exam1}
	\begin{align}
\text{Obj.} \  &	\max_{\mathcal{S}_t \subseteq [N]}\, \mathbb{E}(\sum_{i\in \mathcal{S}_t}   d_{i,t}  z_{i,t}) = \sum_{i\in \mathcal{S}_t}   d_{i,t}  p_{i,t} \label{eq:exam1:obj}\\ \ 
\text{s.t.} \ & \  \sum_{i\in \mathcal{S}_t} r_{i,t}  \leq b_t \label{eq:exam1:con}
	\end{align}
\end{subequations}
\qed

\textbf{Example 2.} Model (\ref{eq:exam2}) \cite{online5} aims to track a load reduction target $D_t$, where the objective  (\ref{eq:exam2:obj})  minimizes the expected squared deviation. Equation (\ref{eq:exam2:con}) is a cardinality constraint on $\mathcal{S}_t$, 
which limits the number of selected customers by $b_t$. This can also be interpreted as the case with unit revenue $r_{i,t}$.
\begin{subequations} \label{eq:exam2}
	\begin{align}
\text{Obj.} \  &	\min_{\mathcal{S}_t \subseteq [N]}\, \mathbb{E}(\sum_{i\in \mathcal{S}_t}   d_{i,t}  z_{i,t} -D_t)^2  \label{eq:exam2:obj}\\ \
\text{s.t.} \ & \ |\mathcal{S}_t| = \sum_{i\in \mathcal{S}_t} 1  \leq b_t \label{eq:exam2:con}
	\end{align}
\end{subequations}

Since $z_{i,t}\sim \text{Bern}(p_{i,t})$ is assumed to be independent across customers, the objective function in (\ref{eq:exam2:obj}) is equivalent to
\begin{align*}
	\min_{\mathcal{S}_t \subseteq [N]}\  (\sum_{i\in \mathcal{S}_t}   d_{i,t}  p_{i,t} -D_t )^2 + \sum_{i\in \mathcal{S}_t} d_{i,t}^2p_{i,t}(1-p_{i,t})
\end{align*}

For the DR cases that positive deviation (more load reduction) from the target is desirable, the objective of (\ref{eq:exam2}) can be modified as
 \begin{align*}
      \text{Obj.} \  &	\min_{\mathcal{S}_t \subseteq [N]}\, \mathbb{E}\left[   
      \max\left \{ D_t-   \sum_{i\in \mathcal{S}_t}  d_{i,t}  z_{i,t} ,\,  0\right\}
      \right]
 \end{align*}
 which aims to only minimize the expected negative deviation from the target.
 \qed

\noindent{\textbf{Remark.} In our problem formulation, the offered credits $r_{i,t}$ and the budget $b_t$ are given as parameters for simplicity. Actually, they can also be optimized to achieve higher DR efficiency from the perspective of LSEs, which relates to the incentive mechanism design  for DR programs \cite{drmech}. }

In the follows, Example 1 with the OCS model (\ref{eq:exam1}) is used for the illustration of algorithm design, 
 while the proposed framework is clearly applicable to other application cases.

\subsection{Contextual MAB Modelling}

If  the  probability profiles $\bm{p}_t:=(p_{i,t})_{i\in[N]}$ of customers are known, the OCS model (\ref{eq:ocs}) is purely an optimization problem.   
However,  the SA barely has access to  $\bm{p}_t$  in practice, which actually depict the customer opt-out behaviors. Moreover, the  probability profiles $\bm{p}_t$ are time-varying and  influenced by various environmental factors. To address this uncertainty issue, the CMAB framework is leveraged to model the residential DR program as an online learning and decision-making problem.  

In the CMAB language, each customer $i\in[N]$ is treated as an independent arm. A set of multiple arms, called a super-arm, constitutes a possible action that the agent takes. At each time $t\in[T]$, the SA  plays the agent role and takes the action $\mathcal{S}_t$. Then the SA observes the outcomes $(z_{i,t})_{i\in \mathcal{S}_t}$ that are generated from the  distributions $\text{Bern}(p_{i,t})$,
and enjoys the reward of   $\sum_{i\in \mathcal{S}_t} d_{i,t}z_{i,t}$. Since the customer opt-out outcome $z_{i,t}$ is binary, the widely-used logistic regression method (\ref{eq:logist}) is utilized to learn the unknown $p_{i,t}$:
\begin{align} \label{eq:logist}
\qquad \ \ p_{i,t} = \frac{\exp(\alpha_i+\bx_{i,t}^\top \bm{\beta}_i)}{1+\exp(\alpha_i+\bx_{i,t}^\top \bm{\beta}_i)}\quad \ \forall i\in[N],t\in[T]
\end{align}
where $\bx_{i,t}\in \mathbb{R}^m$  is the feature vector that captures the environmental factors for customer $i$ at $t$-th DR event. Each entry of $\bm{x}_{i,t}$ corresponds to a quantified factor, such as the offered revenue, indoor temperature, real-time electricity price, the fatigue effect of being repeatedly selected, etc.
 $\bm{\beta}_i\in \mathbb{R}^m$  is the weight vector describing how customer $i$ reacts to those factors, and $\alpha_i$ denotes his individual preference. 
Denote $\hat{\bx}_{i,t}:=(1,\bx_{i,t})$ as the context vector and $\bth_i: = (\alpha_i, \bm{\beta}_{i})$, then the linear term in (\ref{eq:logist}) becomes $\hat{\bx}_{i,t}^\top \bth_i$, and the unknown parameter of each customer $i$ is summarized by $\bth_i$.

As a result, the sequential customer selection in residential DR is modelled as a CMAB problem. The SA aims to learn the unknown $(\bth_i)_{i\in[N]}$ and improves the customer selection strategies. To this end, we propose the online learning and selection (OLS) algorithm to solve this CMAB problem efficiently, which is elaborated in the next section.



\section{Algorithm Design} \label{sec:algorithm}

In this section, the TS algorithm is 
introduced, and  the offline optimization method and Bayesian inference method  are presented. Then 
we assemble these methods and develop the OLS algorithm for residential DR.



\subsection{Thompson Sampling Algorithm}

Consider a general $T$-times MAB problem where an agent selects an action  $a_t$ from the action set $\mathcal{A}_t$ at each time $t$. After applying action $a_t$, the agent observes an outcome $u_t$, which is  randomly generated
from a conditional probability distribution $\mathcal{P}_\theta(\cdot|a_t)$, and then  obtains a reward $R_t = R(u_t)$ with known deterministic  function $R(\cdot)$. The agent intends to maximize the total expected reward but is initially uncertain about the value of $\theta$. 
Thompson sampling (TS) algorithm \cite{ts1} is a Bayesian learning framework that solves such MAB problems with effective balance between exploration and exploitation.

\begin{algorithm}
 \caption{Thompson Sampling Algorithm}
 \begin{algorithmic}[1]  \label{tsalg}
 
   \STATE \textbf{Input:} Prior distribution $\mathcal{P}$ on $\theta$. 
  \FOR {$t = 1$ to $T$}
  
  \STATE Sample $\hat{\theta}\sim \mathcal{P}$.
  \STATE $a_t \leftarrow \arg\max_{a\in \mathcal{A}_t} \mathbb{E}_{\mathcal{P}_{\hat{\theta}}}[R(u_t)|a_t = a] $. 
  
  Apply $a_t$ and observe $u_t$.
  
  \STATE Posterior distribution update:
  \begin{align*}
     \mathcal{P}\leftarrow \frac{\mathcal{P}(\theta) \mathcal{P}_\theta (u_t|a_t)}{\int_{\tilde{\theta}} \mathcal{P}(\tilde{\theta}) \mathcal{P}_{\tilde{\theta} }(u_t|a_t)\, d\tilde{\theta}}\qquad\quad 
  \end{align*}
  \ENDFOR
 \end{algorithmic} 
 \end{algorithm}

As illustrated in Algorithm \ref{tsalg}, TS algorithm represents the initial belief on $\theta$ using a prior distribution $\mathcal{P}$. At each time $t$, TS draws  a random sample $\hat{\theta}$ from $\mathcal{P}$, then takes the optimal action based on the sample $\hat{\theta}$. After the outcome $u_t$ is observed, the Bayesian rule is applied to update the belief and obtain the posterior distribution of $\theta$. 
There are three key observations about the TS algorithm:
\begin{itemize}
  
    \item [1)] As outcome data accumulate, the predefined prior distribution will be washed out and the posterior converges to the true distribution or true value of $\theta$.
    \item [2)] The TS algorithm encourages exploration by the random sampling. As the posterior distribution gradually concentrates, less exploration and more exploitation will be performed, which strikes an effective balance.
    \item [3)] The crucial advantage of TS algorithm is that  the complex online problem is decomposed into a Bayesian learning task and a deterministic optimization task.
    In particular, the optimization problem remains the original model formulation, which enables efficient solution methods.
\end{itemize}

In the follows, the specific methods for the optimization task and the Bayesian learning task are presented respectively.

\subsection{Offline Optimization Method}

At each DR event, 
given the customer probability profiles $\bm{p}_{t}$, the OCS model (\ref{eq:exam1}) can be equivalently reformulated as the  binary optimization problem (\ref{eq:refo}):
\begin{subequations} \label{eq:refo}
	\begin{align}
	\text{Obj.} \  &	\max_{y_{i,t}\in\{0,1\}}\  \sum_{i=1}^N   d_{i,t}  p_{i,t}y_{i,t} \\ 
	\text{s.t.} \ & \sum_{i=1}^N r_{i,t} y_{i,t} \leq b_t
	\end{align}
\end{subequations}
where binary variable $y_{i,t}\in\{0,1\}$ is introduced to indicate  whether the SA selects customer $i$ or not at $t$-th DR event.

The binary optimization model  (\ref{eq:refo}) can be solved  efficiently using many  available optimizers such as IBM CPLEX and Gurobi, which are employed as the offline solution tools. Let $\bm{y}_t^*: = (y_{i,t}^*)_{i\in[N]}$ be the optimal solution of model (\ref{eq:refo}). For concise expression, the optimizer tools are denoted as an offline oracle $
\mathcal{O}:\bm{p}_t\rightarrow \bm{y}_t^*$. It is worth mentioning that
when taking the power flow constraints  (\ref{eq:dist}) (\ref{eq:plcon}) in Appendix \ref{app:powerflow} into consideration, the OCS model becomes a mixed integer linear programming, which can be solved efficiently as well.

\subsection{Bayesian Inference for Logistic Model}\label{sec:bayes}

For clear expression,  we abuse notations a bit and discard subscripts $i$ and $t$ in this part. Under the TS framework, a prior distribution  $\sP(\bth)$ on the unknown parameter 
$\bth$ is constructed.
After  the outcome $(\hat{\bx}, z)$ is observed, the posterior distribution is calculated by the Bayesian law:
\begin{align}\label{eq:bayesian}
    \mathcal{P}(\bth|\hat{\bx}, z) = \frac{\mathcal{P}(\bth)\mathcal{P}(z|\bth,\hat{\bx})}{\int_{\tilde{\bth}}\mathcal{P}(\tilde{\bth})\mathcal{P}(z|\tilde{\bth},\hat{\bx})\, d\tilde{\bth}}
\end{align}
and the logistic likelihood function $\mathcal{P}(z|\bth,\hat{\bx})$ is given by
\begin{align} \label{eq:logis2}
\mathcal{P}(z|\bth,\hat{\bx}) = \phi\left((2z-1)\hat{\bx}^\top \bth\right) 
\end{align}
where $\phi(x):= 1/(1+e^{-x})$ and (\ref{eq:logis2}) is equivalent to (\ref{eq:logist}).

Due to the analytically inconvenient form of the likelihood function, Bayesian inference for the logistic regression model is recognized as an intrinsically hard problem \cite{pg1}, 
thus the exact posterior  $\mathcal{P}(\bth|\hat{\bx}, z)$ (\ref{eq:bayesian}) is intractable to compute. { To address this issue, the concept of conjugate prior \cite{conju} is leveraged to obtain a closed-form expression for the posterior update. Specifically, let the prior be  a Gaussian distribution with $\mathcal{P}(\bth)\sim \mathcal{N}(\bm{\mu},\bm{\Sigma})$. Then the  variational Bayesian inference approach  \cite{varia} is employed to approximate the logistic likelihood function (\ref{eq:logis2}) with a Gaussian-like distribution.}
 The fundamental tool at the heart of this approach  is a lower bound approximation of  (\ref{eq:logis2}):
\begin{align}\label{eq:lowerbound}
       \mathcal{P}(z|\bth,\hat{\bx}) \geq \underbrace{ \phi(\xi) \exp\left[ \frac{s-\xi}{2}+\ell(\xi) (s^2-\xi^2) \right]}_{:= \mathcal{P}(z|\bth,\hat{\bx},\xi)}
\end{align}
where $s: =(2z-1)\hat{\bx}^\top \bth$, $\ell(\xi): = (1/2-\phi(\xi))/2\xi$, and $\xi$ is the variational parameter. 

The variational distribution $\mathcal{P}(z|\bth,\hat{\bx},\xi)$ in (\ref{eq:lowerbound}) has a  convenient property that it depends on  $\bth$
 only quadratically in the exponent. As the prior is a Gaussian distribution with $\mathcal{P}(\bth)\sim \mathcal{N}(\bm{\mu},\bm{\Sigma})$, we use  the Gaussian-like variational distribution  $\mathcal{P}(z|\bth,\hat{\bx},\xi)$ to approximate the logistic likelihood function $ \mathcal{P}(z|\bth,\hat{\bx})$ in the Bayesian inference (\ref{eq:bayesian}). As a result,  
 the posterior is also a Gaussian distribution  $  \mathcal{P}(\bth|\hat{\bx}, z)\sim \mathcal{N}(\mathcal{\hat{\bm{\mu}}}, \hat{\bm{\Sigma}})$ with the closed-form update rule (\ref{eq:update}):
\begin{subequations} \label{eq:update}
    \begin{align}
         \hat{\bm{\Sigma}}^{-1} & = {\bm{\Sigma}}^{-1}+2|\ell(\xi)|  \hat{\bx}\hat{\bx}^\top\\
         \mathcal{\hat{\bm{\mu}}} \ & =  \hat{\bm{\Sigma}}\left[ {\bm{\Sigma}^{-1}}\bmu +(z-\frac{1}{2})\hat{\bx}\right]
    \end{align}
\end{subequations}
 See \cite{varia} for the detailed derivation. 
 
 Since the posterior covariance matrix $\hat{\bm{\Sigma}}$  depends on the variational parameter $\xi$,  its value needs to be specified such that the lower bound approximation in (\ref{eq:lowerbound}) is optimized. The optimal $\xi$ is achieved by   maximizing the expected complete log-likelihood function $\mathbb{E}\left[\log \mathcal{P}(\bth) \mathcal{P}(z|\bth,\hat{\bx},\xi) \right]$, where the expectation is taken over $\mathcal{P}(\bth|\hat{\bx},z,\xi^{\text{old}})$, and this
 leads to a closed form solution:
 \begin{align}\label{eq:xi}
     \xi = \sqrt{\hat{\bx}^\top\hat{\bm{\Sigma}}\hat{\bx} +(\hat{\bx}^\top \hat{\bmu})^2  }
 \end{align}
Alternating between  the posterior update (\ref{eq:update}) and the $\xi$ update (\ref{eq:xi}) 
monotonically improves the posterior approximation \cite{varia}. Generally, by two or three iterations, an accurate approximation can be achieved.

\subsection{Online Learning and Selection Algorithm}

For each customer $i\in [N]$, we construct a Gaussian prior $\sN(\bmu_i, \bm{\Sigma}_i)$ on the unknown $\bth_i$ based on historical information. 
Using the TS framework, the online learning and selection (OLS) algorithm  for the residential DR problem is developed as Algorithm \ref{onlinealg}. From statement \ref{st:sample_op} to \ref{st:sample_ed}, it generates a random sample of $\hat{\bth}_{i,t}$ and then calculates the probability $p_{i,t}$ with the contextual information $\hat{\bx}_{i,t}$ for each customer. With the obtained probability profiles $\bm{p}_t$, the SA determines the optimal selection of customers for load reduction by solving the OCS model (\ref{eq:refo}), when available optimizers can be used. After observing the  behavior outcome  $z_{i,t}$ of each selected customer, statements \ref{st:update_op} - \ref{st:update_ed} update the posterior on $\bth_i$ using the variational Bayesian inference approach in section \ref{sec:bayes}. In Step \ref{st:postupdate}, the alternation between the posterior update and the $\xi_i$ update is performed three times to obtain accurate posterior approximation \cite{varia}.

 \begin{algorithm}
 \caption{ Online Learning and Selection Algorithm}
 \begin{algorithmic}[1]  \label{onlinealg}
 \renewcommand{\algorithmicrequire}{\textbf{Input:}}
 \renewcommand{\algorithmicensure}{\textbf{Output:}}

  \STATE \textbf{Input:} $(\bmu_i, \bm{\Sigma}_i)$ of each customer $i\in[N]$. 
  
  \FOR {$t = 1$ to $T$}
  \STATE Receive $\{d_{i,t}, r_{i,t}, \hat{\bx}_{i,t}\}$ from each customer $i\in[N]$. The SA sets the budget parameter $b_t$.
  \FOR {customer $i = 1$ to $N$ {(in parallel)}} \label{st:sample_op}
    
  \STATE Sample $\hat{\bth}_{i,t}\sim \mathcal{N}\left( {\bm{\mu}}_{i}, {\bm{\Sigma}}_{i}   \right)$.

   \STATE   $\, p_{i,t}  \; \leftarrow \; 1/\left[{1+\exp(-\hat{\bx}_{i,t}^\top \hat{\bth}_{i,t})} \right]$.
  \ENDFOR \label{st:sample_ed}

 \STATE  Solve the OCS problem (\ref{eq:refo}) with oracle $\bm{y}_t\leftarrow \mathcal{O}(\bm{p}_t)$. \label{st:opt}
 
   Reduce loads for those customers with $y_{i,t}=1$, and observe the responses $z_{i,t}$.

 \FOR { customer $i\in [N]$ with $y_{i,t}=1$ } \label{st:update_op}
   \STATE Initialize $\xi_i$ by
$
        \xi_i \leftarrow  \sqrt{\hat{\bx}_{i,t}^\top{\bm{\Sigma}_i}\hat{\bx}_{i,t} +(\hat{\bx}_{i,t}^\top {\bmu}_i)^2  }.
$
   \STATE   Iterate three times between the posterior update \label{st:postupdate}
     \begin{align*}
     \begin{cases}
       \bm{\hat{\Sigma}}_i^{-1}  \leftarrow {\bm{\Sigma}_i}^{-1}+2|\ell(\xi_i)|  \hat{\bx}_{i,t}\hat{\bx}_{i,t}^\top,\\
       \   \bm{\hat{\mu}}_i   \  \, \leftarrow  \bm{\hat{\Sigma}}_i\left[ {\bm{\Sigma}_i^{-1}}\bmu_i +(z_{i,t}-\frac{1}{2})\hat{\bx}_{i,t}\right]
     \end{cases}
   \end{align*}
  and  the $\xi_i$ update 
  \begin{align*}
       \xi_i\ \leftarrow  \sqrt{\hat{\bx}_{i,t}^\top{\bm{\hat{\Sigma}}_i}\hat{\bx}_{i,t} +(\hat{\bx}_{i,t}^\top \bm{\hat{\mu}}_i)^2  }.\ \
  \end{align*}
\STATE Set
 ${\bm{\Sigma}_i} \  \leftarrow   \bm{\hat{\Sigma}}_i,\ \mathcal{{\bm{\mu}}}_i \ \leftarrow \bm{\hat{\mu}}_i.$

  \ENDFOR \label{st:update_ed}
 
  \ENDFOR
 \end{algorithmic} 
 \end{algorithm}

On the one hand, the OLS algorithm inherits the merits of TS, and decomposes the online DR problem into learning and optimization two separate parts. Since the optimization problem remains the original formulation without being corrupted by the learning task, the OLS algorithm can be applied to the practical DR problems with  complicated objectives and constraints. On the other hand, the closed form  formula for posterior update enables very convenient Bayesian inference on the unknown parameter, which leads to efficient implementation of the OLS algorithm.

Besides, preserving customer data privacy is an important issue for the practical implementation of DR programs, which is related to the  fields of data storage, signal process communication, etc. 
There have been a large number of studies \cite{priv1,priv2} devoted to this problem, whose schemes can be applied to the implementation of the proposed OLS algorithm to ensure customer privacy.

\section{Performance Analysis}\label{sec:regret}

In this section, we provide the performance analysis for the proposed online algorithm and prove that it achieves a $O(\sqrt{T}\log T)$ Bayesian regret bound for the OCS problem (\ref{eq:exam1}) when exact Bayesian inference is used.

\subsection{Main Result}
Define $\hat{m}:=m+1$ as the dimension of $\bth_i$.
Let $\bm{\Theta}: = (\bth_i)_{i\in[N]}\in \mathbb{R}^{N\hat{m}}$ collect all the unknown customer parameters. Denote  the reward received at time $t$ as $R_t: = \sum_{i\in \sS_t} d_{i,t}z_{i,t}$
and define the reward function $f^{\bTh}_t$ as
\begin{align}\label{eq:rewardfunc}
    f^{\bTh}_t(\sS_t) = \mathbb{E}(R_t|\bm{\Theta},\sS_t)= \sum_{i\in \sS_t} \frac{d_{i,t}}{1+\exp{(-\hat{\bx}_{i,t}^\top \bth_i)}}
\end{align}

To measure the performance of the online algorithm, we define the $T$-period regret as 
\begin{align} \label{eq:regret}
   \text{Regret}(T,\bTh) = \sum_{t=1}^T \mathbb{E}\left[ f^{\bTh}_t(\sS_t^*) -  f^{\bTh}_t(\sS_t)\, |\,\bTh \right]
\end{align}
where $\sS_t^* \in \arg\max_{\sS_t\in \mathcal{A}_t} f^{\bTh}_t(\sS_t)$ is the optimal solution of model (\ref{eq:exam1}) with known $\bTh$, and $\mathcal{A}_t$ denotes the feasible region described by constraint (\ref{eq:exam1:con}), while $\sS_t$ denoted the customer selection decision made by the online algorithm. The expectation in (\ref{eq:regret}) is taken over the randomness of $\sS_t$. Generally, a sublinear regret over the time length $T$ is desired, which indicates  that the online algorithm can eventually  learn the optimal solution, since $\text{Regret}(T,\bTh)/T\to 0$ as $T\to \infty$.

Since $\bTh$ is treated as a random variable in the TS framework,  we further  define the  $T$-period Bayesian regret as 
\begin{align} \label{eq:BayesRegret}
   \text{BayRegret}(T) = \mathbb{E}_{\bTh\sim \sP(\bTh)}\left[\text{Regret}(T,\bTh)\right]
\end{align}
which  is the expectation of $ \text{Regret}(T,\bTh)$ taken over the prior distribution $\sP(\bTh)$ of $\bTh$. 
It can be shown that asymptotic bounds on Bayesian regret are essentially asymptotic bounds on regret with the same order. See \cite{ts2} for more explanations for Bayesian regret. 

We further make the following two assumptions. Assumption \ref{ass:exact} is generally made for the theoretic analysis of Bayesian regret. In practical applications, performing exact Bayesian inference is intractable for the cases without conjugate prior, thus approximation approaches or Gibbs sampler are usually employed to obtain  posterior samples \cite{ts1}. For example, the variational approximation method introduced in Section \ref{sec:bayes} and the P\'olya-Gamma Gibbs sampler in \cite{pg1} are two typical Bayesian inference approaches for logistic models.

\begin{assumption}\label{ass:exact}
Exact Bayesian inference is performed in the OLS algorithm, i.e., the exact posterior distribution of ${\bTh}$ is obtained and used for sampling at each time $t\in[T]$.
\end{assumption}

\begin{assumption}\label{ass:boundthe}
$\bTh$ is bounded with $||\bTh||_\infty \leq L$.
\end{assumption}

Besides, the context vectors are normalized such that $\hat{\bx}_{i,t}$ is bounded with $\|\hat{\bx}_{i,t}\|_\infty \leq 1$ for all $i\in[N]$ and $t\in[T]$.
We show that the proposed OLS algorithm can achieve a sublinear Bayesian regret with the order of $O(\sqrt{T}\log T)$, which is stated formally as Theorem \ref{main theorem}.

\begin{theorem}\label{main theorem}
Under Assumption \ref{ass:exact} and \ref{ass:boundthe}, 
the Bayesian regret bound  of the OLS algorithm  for the OCS problem (\ref{eq:exam1}) is 
\begin{align*}
    \textnormal{BayRegret}(T)\leq O\left(\bar{D}\hat{m}N^\frac{3}{2}e^{\hat{m}L}{\bar{d}}/{\underline{d}}\cdot\sqrt{T}\log T\right)
\end{align*}
where 
$\bar{d}:=\sup_{i,d} d_{i,t}$, $\underline{d}:= \inf_{i,t} d_{i,t}>0$, and $\bar{D}$  is defined in (\ref{eq:boundf}).
\end{theorem}

The theoretic results in \cite[Section 7]{ts2} are applied to prove Theorem \ref{main theorem} in the next subsection.

\subsection{Proof Sketch of Theorem \ref{main theorem}}

We first show that our problem formulation satisfies the two assumptions imposed in \cite[Section 7]{ts2}. For \cite[Assumption 1]{ts2}, the reward function (\ref{eq:rewardfunc}) is bounded by
\begin{align}\label{eq:boundf}
\qquad\qquad 0\leq   f^{\bTh}_t \leq \sup_{t\in[T]}\sum_{i\in[N]} d_{i,t}:= \bar{D}, \quad \forall t\in[T]
\end{align}
In terms of \cite[Assumption 2]{ts2}, 
we have the following lemma, whose proof is provided in Appendix \ref{pf-lemma-1}.

\begin{lemma}\label{lemma:subgaussian}
For all $t\in[T]$, $R_t - f^{\bTh}_t(\mathcal{S}_t)$ conditioned on $(\bTh, \mathcal{S}_t)$ is $\frac{1}{2}\bar{D}$-sub-Gaussian.
\end{lemma}

Under Assumption \ref{ass:boundthe},  define the reward function class as 
\begin{align}
   \qquad \mathcal{F}_t: =\left\{f_t^{\bTh}\,|\,\bTh\in \Psi\right\}, \ t\in[T]
\end{align}
where $\Psi:=\{\bTh\in \R^{N\hat{m}}\,|\, ||\bTh||_\infty\leq L\}$.
The reward function class $\mathcal{F}_t$ is time dependent  due to the time-variant $d_{i,t}$ and $\hat{\bx}_{i,t}$. This is  a bit different from the setting in \cite[Section 7]{ts2}, but it can be checked the regret analysis still holds since $d_{i,t}$ and $\hat{\bx}_{i,t}$ are given parameters and the task is to learn the unknown $\bTh$. By applying the result in  \cite[Proposition 10]{ts2}, we have the following Bayesian regret bound.
\begin{lemma}\label{lemma: main}
Under Assumption \ref{ass:exact} and \ref{ass:boundthe},  the Bayesian regret of the OLS algorithm is bounded by
\begin{align}\label{eq:main lemma}
\begin{split}
    & \textnormal{BayRegret}(T)\leq \sup_{t\in[T]} \Big\{ 1+\left[\textnormal{dim}_E(\F_t,T^{-1})+1\right]\bar{D}  \\ &
    +8\bar{D}\sqrt{\textnormal{dim}_E(\F_t,T^{-1})(1+o(1)+\textnormal{dim}_K(\F_t))T\log T}\, \Big\}
\end{split}
\end{align}
\end{lemma}
In (\ref{eq:main lemma}),  $\textnormal{dim}_K(\F_t)$ is the Kolmogorov dimension defined by \cite[Definition 1]{ts2} and $\textnormal{dim}_E(\F_t,T^{-1})$ denotes the ${1}/{T}$-eluder dimension defined by \cite[Definition 3]{ts2} of function class $\F_t$. 
Intuitively, the Kolmogorov dimension is related to the measure of complexity to learn a function class, while the eluder dimension captures how effectively the unknown values can be inferred from the observed samples.
To achieve the final result, we further bound $\textnormal{dim}_E(\F_t,T^{-1})$ and $\textnormal{dim}_K(\F_t)$ in (\ref{eq:main lemma}) by the following two lemmas, whose proofs are provided in Appendix \ref{pf-lemma-3} and \ref{pf-lemma-4} respectively.

\begin{lemma}\label{le:E-dim}
Under Assumption \ref{ass:boundthe}, we have
\begin{align}\label{eq:e-dimbound}
  \sup_{t\in [T]}  \textnormal{dim}_E(\F_t,T^{-1})\leq C_3\log\left[C_2(1+C_1 T^2)\right]+1
\end{align}
where the definitions of constants $C_1, C_2, C_3$ are given as (\ref{eq:Cdef}).
\end{lemma}

\begin{lemma}\label{le:K-dim}
Under Assumption \ref{ass:boundthe}, we have
\begin{align}\label{eq:k-dimbound}
     \sup_{t\in [T]}  \textnormal{dim}_K(\F_t)\leq N\hat{m}
\end{align}
\end{lemma}

Combing the results in Lemma \ref{lemma: main}-\ref{le:K-dim}, we obtain Theorem \ref{main theorem}.

\hfill $\Box$


\section{Numerical Simulations} \label{sec:simulation}

\subsection{Simulation Setting}

Consider the residential DR with $N=1000$ customers.  
The reduced load $d_{i,t}$ and the offered revenue $r_{i,t}$ are randomly and independently generated from the uniform distribution $\text{Unif}\,[0,1]$ for each customer, which are fixed for different time steps. At each time $t$, the SA randomly samples a budget  $b_t$ from $\text{Unif}\,[300,400]$ to recruit customers for DR participation.
Set the number of contextual features as $m=9$, and let 
all customers share a same  feature vector ${\bx}_{t}\in \mathbb{R}^{9}$ at each time $t$, whose elements are
independently generated from  $\text{Unif}\,[0,2]$. 
Assume that there exists an underlying ground truth $\bm{\theta}_i^*\in \mathbb{R}^{10}$ associated with each customer, which is generated from $\text{Unif}[-0.4,0.6]$ to simulate the customer behaviors. 
In the OLS algorithm, we assign a Gaussian prior distribution $\mathcal{N}(\bm{\theta}_i^*+\delta\bm{u}_i, \sigma^2\bm{I} )$ for each customer, where each element of $\bm{u}_i$ is  randomly generated from $\text{Unif}[-1,1]$ and $\bm{I}$ denotes the identity matrix.
Parameters $\delta$ and $\sigma$ denote the mean error and standard deviation level of the prior distribution respectively.

 For the OLS algorithm,  it takes  0.27 second on average to solve the OCS problem (\ref{eq:refo})\footnote{The simulations are  performed in a  computing environment with Intel(R) Core(TM) i7-7660U CPUs running at 2.50 GHz and with 8-GB RAM. 
All the programmings are implemented in Matlab 2018b. We use the CVX package to model the binary optimization program (\ref{eq:refo}), and Gurobi   7.52 is employed as the solver. }, 
     and averagely 0.4 second  to finish one entire round including the optimization task and the learning task. 

\subsection{Algorithm Performance}\label{sec:sim:com}

We compare the proposed OLS algorithm with the UCB-based online learning DR algorithm proposed in \cite{online5}, which does not consider the contextual influence on customer behaviors. We set $\delta = 0.3, \sigma =0.3$ for the prior distributions in the OLS algorithm, and let the used UCB function be 
\begin{align*}
p_{i,t} = \bar{p}_{i,t} + \sqrt{   \frac{3\log t}{2 T_i(t)}}
\end{align*}
where $\bar{p}_{i,t}$ is the sample average of the historically realized $z_{i,t}$ by time $t$, and $T_i(t)$ denotes the number of times selecting customer $i$ by time $t$. 

\begin{figure}[thpb]
	\centering
	\includegraphics[scale=0.262]{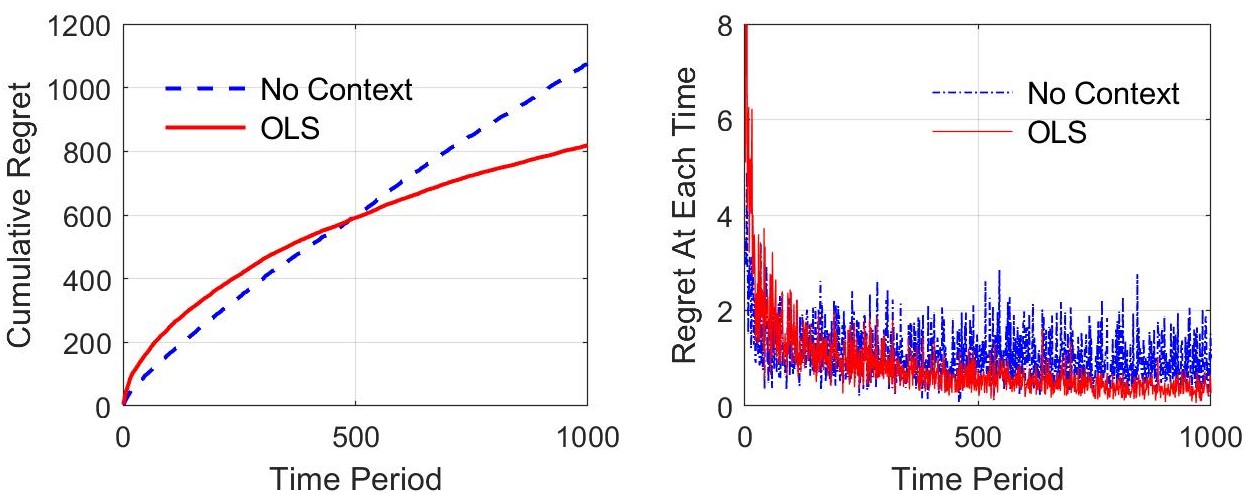}
	\caption{Regret comparison between the OLS algorithm and the UCB-based online learning DR algorithm without contexts.}
	\label{fig:tsucb}
\end{figure}

The simulation results are shown as Figure  \ref{fig:tsucb}. It is observed that the OLS algorithm exhibits a sublinear cumulative regret curve, and its regret at each time gradually decreases to zero value. In contrast, without considering the contextual factors, the UCB-based DR algorithm does not learn the customer behaviors well and maintains high regret at each time.


\begin{figure}[thpb]
	\centering
	\includegraphics[scale=0.225
	]{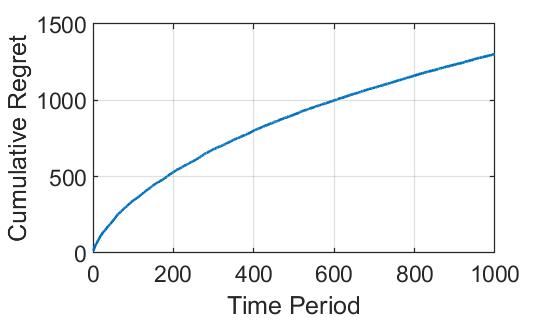}
	\includegraphics[scale=0.225]{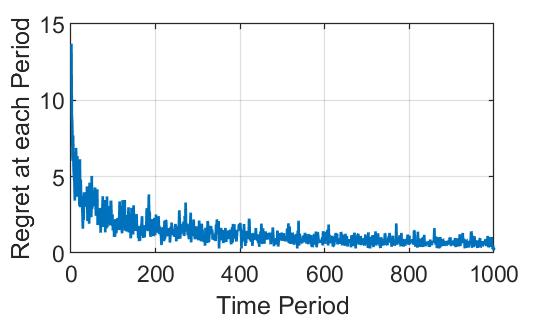}
	\caption{Regret results of the OLS algorithm with fixed credit $r_{i,t}=0.5$.}
	\label{fig:fixrew}
\end{figure}

To test the algorithm performance with fixed incentives, we set $r_{i,t}$ as 0.5 for all customers and times. The regret results are shown as the  Figure \ref{fig:fixrew}.  From the simulation results, it is seen that the cumulative regret exhibits a sublinear trend over time periods and the regret at each period decreases quickly. The OLS algorithm performs well for the case with fixed incentives. 

From the simulation results, it is observed that  the regret at each time decreases dramatically within the first 100 periods, which exhibits the benefits and efficiency of the proposed method. Moreover, historical data can be leveraged to start the learning process with  a good initial  model. Some 
    techniques can be used to accelerate the customer learning  as well. 
For  example, we can cluster customers into multiple groups according to their  socio-demographic characteristics and DR behaviors, then combine and share the sample data among each group, which  may be able to speed up the learning process.

\begin{figure}[thpb]
	\centering
	\includegraphics[scale=0.17]{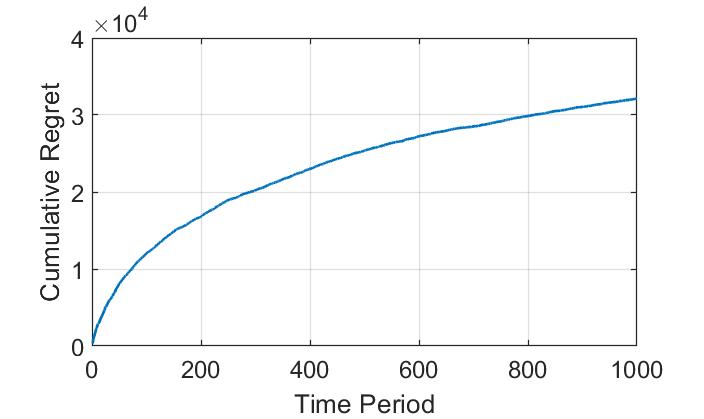}
	\includegraphics[scale=0.17]{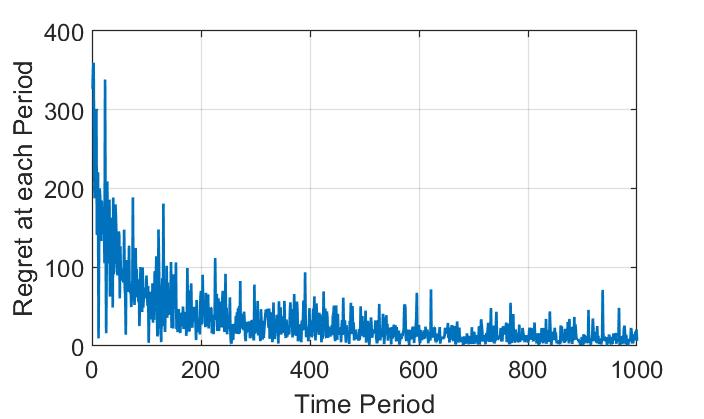}
	\caption{Regret results of the OLS algorithm on the OCS problem (\ref{eq:exam2}).}
	\label{fig:prob3}
\end{figure}

Moreover, we test the algorithm performance on the OCS problem (\ref{eq:exam2}), which aims to track a given load reduction trajectory. The  load reduction target $D_t$ is randomly generated from $\text{Unif}[200,300]$ at each time.
The simulation results are shown as Figure \ref{fig:prob3}, where a similar sublinear cumulative regret is observed.

\subsection{Effects of Prior Distribution}\label{sec:sim:analy}

\begin{figure}[thpb]
	\centering
	\includegraphics[scale=0.26]{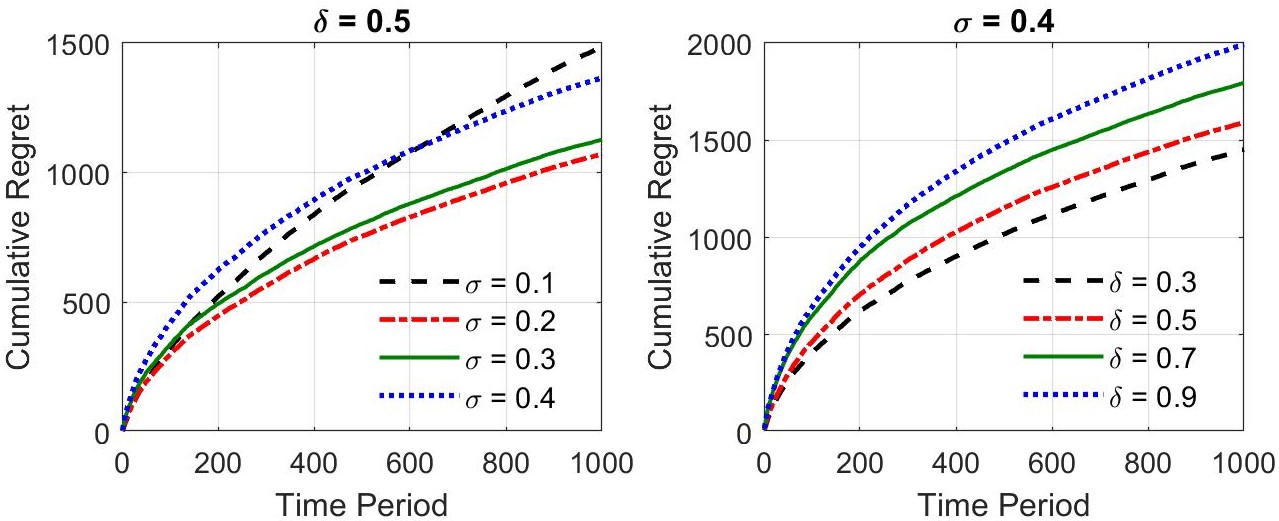}
	\caption{Cumulative regret under different $\delta$ and $\sigma$. (Left: fix $\delta=0.5$, tune $\sigma$ from 0.1 to 0.4. Right: fix $\sigma =0.4$, tune $\delta$ from 0.3 to 0.9.)}
	\label{fig:varmean}
\end{figure}

This part studies the effects of the prior distributions and tests the proposed OLS algorithm with  different parameters $\delta$ and $\sigma$. The simulated regret results are illustrated as Figure \ref{fig:varmean}.  It is observed that when the mean error $\delta$ is fixed, the cumulative regret is generally higher with a larger standard deviation $\sigma$, because a larger $\sigma$ indicates greater uncertainty about the true value and leads to more explorations. However, if the 
standard deviation $\sigma$ is too small, e.g. the case with $\sigma =0.1$, a high cumulative regret occurs, because it sticks to the erroneous prior belief and does not perform sufficient explorations. As for the case with fixed standard deviation  $\sigma=0.4$,  it is seen that the cumulative regret becomes lower as the mean error $\delta$ decreases, which is consistent with the intuition that a more accurate prior belief leads to better performance. 


\section{Conclusion}\label{sec:conclusion}

In this paper, the contextual MAB method is employed to model the customer selection problem in residential DR, considering the uncertain
customer behaviors and the influence of contextual factors. Based on TS framework, the OLS algorithm is developed to learn customer behaviors and select appropriate customers for load reduction with the balance between exploration and exploitation. The simulation results demonstrate  the necessity to consider the contextual factors and 
the learning effectiveness of the proposed  algorithm. For future work, there are two attempt directions: 1) develop the optimal real-time control schemes for  load devices during Phase 1 of the DR event, considering physical system dynamics;
2) study how to design the incentive mechanisms that optimize the  credits  and 
budget  to achieve higher DR efficiency, together with the learning process..

\appendices

\section{Power Flow Constraints}\label{app:powerflow}

Consider an underlying power distribution network delineated by the graph $G(\mathcal{V},\mathcal{E})$, where $\mathcal{V}$ denotes the set of buses and $\mathcal{E}\subset \mathcal{V}\times \mathcal{V}$ denotes the set of distribution lines. For line $jk\in \mathcal{E}$,
denote $P_{jk,t}$ and $Q_{jk,t}$ as the active and reactive power flow  from bus $j$ to bus $k$ at time $t$;
and denote $R_{jk}$ and $X_{jk}$ as the line resistance and reactance. For bus $k\in\mathcal{V}$, denote $P^{in}_{k,t}$ and $Q^{in}_{k,t}$ as the net active and reactive power injection right before the $t$-th DR event, and let $U_{k,t}$ be the squared voltage magnitude of bus $k$.
Using the linearized Distflow model \cite{dist}, the power flow equations are formulated as (\ref{eq:dist}):
\begin{subequations}\label{eq:dist}
\begin{align}
  \sum_{ l:k\to l} P_{kl,t} -  \sum_{j:j\to k} P_{jk,t} &= P_{k,t}^{in} + \sum_{i\in C(k)}y_{i,t}d_{i,t} \\
   \sum_{ l:k\to l} Q_{kl,t} -  \sum_{j:j\to k} Q_{jk,t} &= Q_{k,t}^{in} + \sum_{i\in C(k)}y_{i,t}\eta_id_{i,t}\label{eq:dist:Q}\\
   U_{j,t} -U_{k,t} = 2(&R_{jk}P_{jk,t}+X_{jk}Q_{jk,t})
\end{align}
\end{subequations}
where $d_{i,t}$ is the active load power of customer $i$ that can be reduced at time $t$. 
Binary variable $y_{i,t}\in\{0,1\}$ denotes whether customer $i$ is selected for load reduction, and $C(k)$ denotes the set of customers whose loads are attached to bus $k$. In (\ref{eq:dist:Q}), a constant load power factor is assumed with the constant $\eta_i$. Then the line thermal constraints and voltage limits are formulated as  (\ref{eq:plcon}):
\begin{subequations}\label{eq:plcon}
\begin{gather}
      P_{jk,t}^2 + Q_{jk,t}^2\leq \bar{S}_{jk}^2, \quad \forall jk \in\mathcal{E}\\
   \underline{U}_k \leq  U_{k,t} \leq \bar{U}_k,\qquad \forall k\in \mathcal{V}
\end{gather}
\end{subequations}
where $\bar{S}_{jk}$ is the apparent power capacity of line $jk$; $\underline{U}_k$ and $\bar{U}_k$ are the lower and upper limits of the squared voltage magnitude respectively.

As a result, the feasible set $\Pi_t$ that captures the 
network and power flow constraints is constructed as (\ref{eq:feasi})
\begin{align}\label{eq:feasi}
 \Pi_t : = \{  \mathcal{S}_{\bm{y}_t}\subseteq [N]\,|\, \text{Equation } (\ref{eq:dist}) (\ref{eq:plcon})     \}
\end{align}
where $\mathcal{S}_{\bm{y}_t}$ denotes the subset of customers with
\begin{align*}
    \begin{cases}
    i\in \mathcal{S}_{\bm{y}_t}, & \text{if }y_{i,t}=1\\
    i\notin \mathcal{S}_{\bm{y}_t}, & \text{if }y_{i,t}=0
    \end{cases} \qquad \forall i\in[N].
\end{align*}

\section{Proof of Lemmas}

\subsection{Proof of Lemma \ref{lemma:subgaussian}} \label{pf-lemma-1}
By definition, $R_t-f^{\bTh}_t(\sS_t) = \sum_{i\in\sS_t}d_{i,t}(z_{i,t}-p_{i,t})$. As $z_{i,t}$ is a Bernoulli random variable with $z_{i,t}\sim \text{Bern}(p_{i,t})$, $z_{i,t}-p_{i,t}$ is $\frac{1}{2}$-sub-Gaussian by Hoeffding's lemma. Then for any $\lambda$,
\begin{align*}
    &\ \E\left[\exp(\lambda\sum_{i\in \sS_t}d_{i,t}(z_{i,t}-p_{i,t}))\right]\\
    =& \prod_{i\in \sS_t}\E \left[\exp(\lambda d_{i,t}(z_{i,t}-p_{i,t}))\right]
    \leq\prod_{i\in \sS_t}\exp(\frac{1}{8}\lambda^2d_{i,t}^2)\\
    =&\exp(\frac{1}{8}\lambda^2\sum_{i\in \sS_t}d_{i,t}^2)
    \leq  \exp(\frac{1}{8}\lambda^2\bar{D}^2)
\end{align*}
where the first inequality is because $z_{i,t}-p_{i,t}$ is $\frac{1}{2}$-sub-Gaussian, and the second inequality is due to  (\ref{eq:boundf}).
Hence, $R_t-f^{\bTh}_t(\sS_t)$ conditioned on $(\bTh,\sS_t)$ is $\frac{1}{2}\bar{D}$ -sub-Gaussian. \hfill $\Box$

\subsection{Proof of Lemma \ref{le:E-dim}}\label{pf-lemma-3}
For any $t\in[T]$ and $\sS_t$, define the $N\hat{m}$-dimension vector
\begin{align*}
    \bm{\chi}_t(\sS_t)=(\hat{\bx}_{1,t}^\top\mathbbm{1} \{1\in \sS_t\},\cdots,\hat{\bx}_{N,t}^\top\mathbbm{1} \{N\in \sS_t\})^\top
\end{align*}
where $\mathbbm{1} \{\cdot\}$ is the indicator function. To facilitate the proof, we abuse the notation a bit and adjust $\bTh$  to the matrix form
\begin{align*}
    \bTh = \text{blockdiag}(\bth_1,\bth_2\cdots,\bth_N  )\in \R^{N\hat{m}\times N}
\end{align*}
Define function $ g(\bm{y})=\sum\limits_{i=1}^
    N d_{i,t}\phi(y_i)  $ for $\bm{y}\in \R^N$ with  $\phi(\cdot)$ defined in (\ref{eq:logis2}).
Define a new reward function class by
\begin{align}\label{eq:newf}
    \hat{\F}_t:=\left\{\hat{f}_t^{\bTh}(\sS_t)=g\left(\bTh^\top\bm{\chi}_t(\sS_t)\right)|\,\bTh\in \Psi\right\}
\end{align}
Denote $ \underline{d}$ and $\bar{d}$ as the lower and upper bound of $d_{i,t}$ for all $i\in [N]$ and $t\in [T] $ respectively with $0<  \underline{d}\leq d_{i,t}\leq \bar{d}$.

Let $\by_t:= \bTh^\top\bm{\chi}_t$, then the partial derivative of $g(\by_t)$ in (\ref{eq:newf}) is lower and upper bounded by
\begin{align*}
\qquad \frac{\underline{d}}{4e^{L\hat{m}}}\leq \frac{\partial g}{\partial y_{i,t}}=\frac{d_{i,t}}{e^{-y_{i,t}}+e^{y_{i,t}}+2}\leq \frac{\bar{d}}{4}, \ \ \forall i\in [N]
\end{align*}
 Note that $\|\bTh\|_2\leq L\sqrt{N\hat{m}}$ and $\|\bm{\chi}_t\|_2\leq \sqrt{N\hat{m}}$. Then by \cite[Proposition 4]{pg4}, we have
 \begin{align}
     \text{dim}_E(\hat{\F}_t,T^{-1})\leq C_3\log\left[C_2(1+C_1 T^2)\right]+1
 \end{align}
 where the constants are given by
 \begin{subequations}\label{eq:Cdef}
  \begin{align}
      C_1&:=4L^2N^2{\hat{m}}^2\\
 C_2&:=(\overline{d}/\underline{d})^2(4N-2)e^{2L\hat{m}}+1\\
     C_3&:=N\hat{m}\frac{e}{e-1}C_2
 \end{align}
 \end{subequations}

Since the relation between the new reward function in (\ref{eq:newf}) and the reward function (\ref{eq:rewardfunc}) is
$
    \hat{f}_t^{\bTh}(\sS_t)
    =f_t^{\bTh}(\sS_t)+\frac{1}{2}\sum\limits_{i\notin \sS_t}d_{i,t}
$, for any $\bTh^1, \bTh^2\in\Psi$, $t\in [T]$ and action $\sS_t$,
we have 
$$f_t^{\bTh^1}(\sS_t)-f_t^{\bTh^2}(\sS_t)=\hat{f}_t^{\bTh^1}(\sS_t)-\hat{f}_t^{\bTh^2}(\sS_t).$$  

Hence, by the definition of eluder dimension in \cite[Definition 3]{ts2}, we obtain $\text{dim}_E(\F_t,{1}/{T})=\text{dim}_E(\hat{\F}_t,{1}/{T})$ for all $t\in[T]$, which leads to Lemma \ref{le:E-dim}.\hfill $\Box$

\subsection{Proof of lemma \ref{le:K-dim}}\label{pf-lemma-4}

For any $t\in[T]$ and $\bTh^1,\bTh^2\in\Psi$, with  function $\phi(\cdot)$ defined in (\ref{eq:logis2}), we have
\begin{equation*}
\begin{split}
   & |f_t^{\bTh^1}(\sS_t)-f_t^{\bTh^2}(\sS_t)|
   \leq\sum_{i\in \sS_t}d_{i,t}|\phi(\hat{\bx}_{i,t}^\top\bth_i^1)-\phi(\hat{\bx}_{i,t}^\top\bth_i^2)|\\
   =&\ \sum_{i\in \sS_t} d_{i,t}|\phi'(\xi_{i,t})||\hat{\bx}_{i,t}^\top(\bth_i^1-\bth_i^2)|\\
   \leq & \ \frac{\hat{m}}{4}\sum_{i\in [N]} d_{i,t}\|\bth_i^1-\bth_i^2\|_{\infty}
   \leq \frac{\hat{m}}{4}\bar{D}\|\bTh^1-\bTh^2\|_{\infty}
\end{split}
\end{equation*}
where the first equality is due to the  Lagrange’s mean value theorem for some appropriate $\xi_{i,t}$, and the second inequality uses the property $|\phi'(\cdot)|\leq \frac{1}{4}$.

Hence, for any $f_t^{\bTh^1}$, $f_t^{\bTh^2} \in \mathcal{F}_t$, we have
\begin{align*}
    \|f_t^{\bTh^1}-f_t^{\bTh^2}\|_{\infty}\leq\frac{\hat{m}}{4}\bar{D}\|\bTh^1-\bTh^2\|_{\infty}
\end{align*}
Thus an $\alpha$-covering of $\mathcal{F}_t$ can be achieved through a $(\frac{4\alpha}{\bar{D}\hat{m}})$-covering of the set $\Psi$. Denote $n_t(\alpha)$ as the $\alpha$-covering number of $\F_t$ in the $\infty$-norm. By the definition of Kolmogorov dimension  \cite[Definition 1]{ts2}, it obtains 
\begin{align*}
    \dim_K(\F_t)& =\limsup\limits_{\alpha\downarrow 0}\frac{\log n_t(\alpha)}{\log(1/\alpha)}\\
    &\leq \limsup\limits_{\alpha\downarrow 0}\frac{\log ( \frac{L\bar{D}\hat{m}}{2\alpha}  )^{N\hat{m}}}{\log(1/\alpha)}= N\hat{m}
\end{align*}
where the inequality is because an $\varepsilon$-covering of the set $\Psi$  requires at most $(\frac{2L}{\varepsilon})^{N\hat{m}}$ elements \cite{covering}. Since the above upper bound is independent of time $t$, Lemma \ref{le:K-dim} is proved.   \hfill $\Box$



\end{document}